\documentclass[12pt]{article}

\begin{document}

\noindent {\small CITUSC/00-018\hfill \hfill
hep-th/0106021\newline NSF-ITP-01-55\hfill }

{\vskip1.5cm}

\begin{center}
{\Large \textbf{2T-Physics 2001}}{\footnote{%
This paper is a summary of lectures delivered at the XXXVII
Karpacz Winter School of Theoretical Physics (Feb. 2001), the
M-theory workshop at the ITP, Santa Barbara (May 2001), and the
10$^{th}$ Tohwa International Symposium on String, Fukuoka, Japan
(July 2001).}}

\bigskip

{\vskip0.5cm}

\textbf{Itzhak Bars}

{\vskip0.5cm}

\textsl{CIT-USC Center for Theoretical Physics \& Department of
Physics and Astronomy}

\textsl{University of Southern California,\ Los Angeles, CA
90089-2535, USA}

{\vskip0.5cm}

\textbf{Abstract}
\end{center}

The physics that is traditionally formulated in one--time-physics
(1T-physics) can also be formulated in two-time-physics
(2T-physics). The physical phenomena in 1T or 2T physics are not
different, but the spacetime formalism used to describe them is.
The 2T description involves two extra dimensions (one time and one
space), is more symmetric, and makes manifest many hidden features
of 1T-physics. One such hidden feature is that families of
apparently different 1T-dynamical systems in $d$ dimensions
holographically describe the same 2T system in $d+2$ dimensions.
In 2T-physics there are two timelike dimensions, but there is also
a crucial gauge symmetry that thins out spacetime, thus making
2T-physics effectively equivalent to 1T-physics. The gauge
symmetry is also responsible for ensuring causality and unitarity
in a spacetime with two timelike dimensions. What is gained
through 2T-physics is a unification of diverse 1T dynamics by
making manifest hidden symmetries and relationships among them.
Such symmetries and relationships is the evidence for the presence
of the underlying higher dimensional spacetime structure.
2T-physics could be viewed as a device for gaining a better
understanding of 1T-physics, but beyond this, 2T-physics offers
new vistas in the search of the unified theory while raising deep
questions about the meaning of spacetime. In these lectures, the
recent developments in the gauge field theory formulation of
2T-physics will be described after a brief review of the results
obtained so far in the worldline approach.

\newpage














\section{Worldline approach}

A crucial element in the formulation of 2T-physics \cite{survey2T}-\cite
{NCu11} is an Sp$\left( 2,R\right) $ gauge symmetry in phase space. All new
phenomena in 2T-physics (including the two times) can be traced to the
presence of this gauge symmetry and its generalizations. In the space of all
worldline theories (i.e. all possible background fields) there is an
additional symmetry that corresponds to all canonical transformations in the
phase space of a particle. After describing the role of these two symmetries
in the worldline formalism we will discuss field theory. In field theory
these two symmetries combine and get promoted to noncommutative U$_{\star
}\left( 1,1\right) $ acting on fields as functions of noncommutative phase
space. U$_{\star }\left( 1,1\right) $ symmetry provides the foundation of
2T-physics in field theory, and leads to a unification of various gauge
principles in ordinary field theory, including Maxwell, Einstein and
high-spin gauge principles.

\subsection{Spinless particle and interactions with backgrounds}

An elementary approach for understanding 2T-physics is offered by
the worldline description of a spinless particle and its
interactions. The action
\footnote{%
This action is a generalization of the familiar elementary
worldline action $ \int d\tau \left[
\dot{X}^{M}P_{M}-{\frac{1}{2}}e\left( \tau \right) \,P^{2}
\right] $ for a free massless particle, as seen by specializing to $%
A_{11}=A_{12}=0,$ $A_{22}=e$ and $Q_{22}=P^{2}.$} has the form \cite{emgrav}
\cite{highspin}
\begin{equation}
I_{Q}=\int d\tau \left[ \dot{X}^{M}P_{M}-{\frac{1}{2}}A^{ij}\left( \tau
\right) \,Q_{ij}(X,P)\right] ,
\end{equation}
where the symmetric $A_{ij}=A_{ji}$ for $i=1,2,$ denotes three
Sp$\left( 2,R\right) $ gauge fields, and the symmetric
$Q_{ij}=Q_{ji}$ are three sp$ \left( 2,R\right) $ generators
constructed from the phase space of the particle on the worldline
$\left( X^{M}\left( \tau \right) ,P_{M}\left( \tau \right) \right)
$. An expansion of $Q_{ij}(X,P)$ in powers of $P_{M}$ in some
local domain, $Q_{ij}(X,P)=\sum_{s}\left( f_{ij}\left( X\right)
\right) ^{M_{1}\cdots M_{s}}P_{M_{1}}\cdots P_{M_{s}},$ defines
all the possible background fields in configuration space $\left(
f_{ij}\left( X\right) \right) ^{M_{1}\cdots M_{s}}$ that the
particle can interact with. The local sp$\left( 2,R\right) $ gauge
transformations are
\begin{equation}
\delta X^{M}=-\omega ^{ij}\left( \tau \right) \frac{\partial
Q_{ij}}{\partial P_{M}},\quad \delta P_{M}=\omega ^{ij}\left( \tau
\right) \frac{\partial Q_{ij}}{\partial X^{M}},\quad \delta
A^{ij}=\partial _{\tau }\omega ^{ij}\left( \tau \right) +\left[
A,\omega \left( \tau \right) \right] ^{ij}. \label{deltaXP}
\end{equation}
The action $I_{Q}$ is gauge invariant, with local parameters $\omega
^{ij}\left( \tau \right) ,$ provided the $Q_{ij}\left( X,P\right) $ satisfy
the sp$\left( 2,R\right) $ Lie algebra under Poisson brackets. This is
equivalent to a set of differential equations that must be satisfied by the
background fields $\left( f_{ij}\left( X\right) \right) ^{M_{1}\cdots M_{s}}$
\cite{emgrav}\cite{highspin}. The simplest solution is the free case denoted
by $Q_{ij}=q_{ij}$ (no background fields, only the flat metric $\eta _{MN}$)
\begin{equation}
q_{ij}=X_{i}^{M}X_{j}^{N}\eta _{MN}:\quad q_{11}=X\cdot X,\quad
q_{12}=X\cdot P,\quad \ \ q_{22}=P\cdot P.  \label{free}
\end{equation}
In the flat case, we have defined $X^M_i$ with only upper
spacetime indices, such that $X^M_1=X^M$ and $X^M_2=P^M$. In the
curved case $P_M$ is always defined with a lower spacetime index.
The general solution with $d+2$ dimensional background fields (see
Eqs.(\ref{bkg1}-\ref{hologr}) below) describes all interactions
of the spinless particle with arbitrary electromagnetic,
gravitational and higher spin gauge fields in $d$ dimensions
\cite{highspin}.

Two timelike dimensions is not an input, it is a result of the
gauge symmetry Sp$(2,R)$ on phase space $\left( X^{M},P_{M}\right)
.$ This gauge symmetry imposes the constraints $Q_{ij}(X,P)=0$ on
phase space as a result of the equations of motion of the gauge
field $A_{ij}\left( \tau \right) $. The meaning of the constraints
is that the physical subspace of phase space should be gauge
invariant under Sp$\left( 2,R\right) .$ There is non-trivial
content in such phase space provided spacetime has $d+2$
dimensions, including two timelike dimensions \cite{old2T}. The
solution of the constraints is a physical phase space in two less
dimensions, that is $ \left( d-1\right) $ spacelike and $1$
timelike dimensions. The non-trivial aspect is that there are many
ways of embedding $d$ dimensional phase space in a given $d+2$
dimensional phase space while satisfying the gauge invariance
constraints. Each $d$ dimensional solution represents a different
dynamical system in 1T-physics, but all $d$ dimensional solutions
holographically represent the same higher dimensional $d+2$ theory
in 2T-physics \cite{old2T}.

The Sp$\left( 2,R\right) $ gauge symmetry is responsible for the
effective holographic reduction of the $d+2$ dimensional spacetime
to (a collection of) $d$ dimensional spacetimes with one-time.
Evidently, in the worldline formalism there is a single proper
time $\tau ,$ but the particle position $ X^{M}\left( \tau \right)
$ has two timelike dimensions $X^{0}\left( \tau \right)
,X^{0^{\prime }}\left( \tau \right) $ and the particle momentum $
P_{M}\left( \tau \right) $ has the corresponding two timelike
components, $ P_{0}\left( \tau \right) ,P_{0^{\prime }}\left( \tau
\right) $. Sp$\left( 2,R\right) $ has 3 gauge parameters and 3
constraints. Two gauge parameters and two constraints can be used
to eliminate one timelike and one spacelike dimensions from the
coordinates and momenta. The third gauge parameter and constraint
are equivalent to those associated with $\tau $ reparametrization,
which is familiar in the 1T-physics worldline formulation (as in
the footnote). In making the three gauge choices one must ask
which combination of the two timelike dimensions is identified
with $\tau ?$ Evidently there are many possibilities, and once
this choice is made, the 1T-Hamiltonian of the system, which will
emerge from the solution of the constraints, will be the canonical
momentum that is conjugate to this gauge choice of time. The many
gauge choices correspond to different looking Hamiltonians with
different 1T dynamical content. In this way, by various gauge
fixing, the same 2T system (with the same fixed set of background
fields) can be made to look like diverse 1T systems. Each 1T
system in $d$ dimensions holographically captures all the
information of the 2T system in $ d+2$ dimensions. Therefore there
is a family of 1T systems that are in some sense dual to each
other. Explicit examples of this holography/duality have been
produced in the simplest 2T model \cite{old2T}, namely the free 2T
particle in $d+2$ dimensions. The free 2T particle of
Eq.(\ref{free}) produces the following 1T holographic pictures in
$d$ dimensions: massless relativistic particle, massive
relativistic particle, massive non-relativistic particle, particle
in anti-de-Sitter space $AdS_{d},$ particle in AdS$_{d-k}\times
$S$^{k}$ space for all $k\leq d-2,$ non-relativistic Hydrogen atom
(1/$r$ potential), non-relativistic harmonic oscillator in one
less dimension, some examples of black holes (for d=2,3), and more
...

The higher symmetry of the $d+2$ system is present in all of the
$d$ dimensional holographic pictures. This symmetry is a global
symmetry that commutes with Sp$\left( 2,R\right) ,$ and therefore
is gauge invariant. Therefore it is a symmetry of the action (not
of the Hamiltonian) for any choice of gauge. Before making a gauge
choice the symmetry is realized in $ d+2$ dimensions. After making
a gauge choice the same symmetry is non-linearly realized on the
fewer $d$ dimensions, and therefore it is harder to detect in many
1T-dynamical systems, although it is present. A special example is
provided by the free 2T particle of Eq.(\ref{free}) which
evidently has a linearly realized SO$\left( d,2\right) $ Lorentz
symmetry. This symmetry is interpreted in various ways from the
point of view of 1T dynamics in the $d$ dimensional holographic
pictures: conformal symmetry for the free massless relativistic
particle, dynamical SO$\left( d,2\right) $ symmetry for the
H-atom, SO$\left( d,2\right) $ symmetry for the particle in
AdS$_{d}$ (n.b. larger than SO$\left( d-1,2\right) $), etc.. The
first two cases were familiar, although they were not usually
thought of as having a relation to higher dimensions. All the
other holographic cases mentioned above, including the
non-relativistic massive particle, harmonic oscillator,
AdS$_{d-k}\times $S$^{k},$ etc. all have the same hidden SO$\left(
d,2\right) $ symmetry which was understood for the first time in
the context of 2T-physics. The generators of the symmetry have
been explicitly constructed for all the cases mentioned
\cite{old2T}. What is more, the symmetry is realized in the same
unitary representation as characterized by the eigenvalues of the
Casimir operators of SO$\left( d,2\right) $. In the classical
theory, in which orders of phase space quantum operators are
neglected, all the Casimir operators for SO$\left( d,2\right) $
seem to vanish (this is a non-trivial representation for the
non-compact group). However, when quantum ordering is taken into
account, the Casimir eigenvalues do not vanish, but take on some
special values corresponding to a special representation. The
ordering of phase space operators is in general difficult, but it
can be implemented explicitly in a few cases (conformal, H-atom,
AdS$_{d-k}\times $S$^{k},$ harmonic oscillator). For all these
cases the quantum quadratic Casimir operator is the same
$C_{2}\left( SO\left( d,2\right) \right) =1-d^{2}/4$, and
similarly one obtains some fixed number for all higher Casimir
operators. This unitary representation is the singleton/doubleton
representation (name depends on $d$). This very specific
representation, which is common to all the 1T dynamical models
mentioned, corresponds to the free 2T particle. This fact is
already part of the evidence of the duality that points to the
existence of the unifying $ d+2 $ or 2T structure underlying these
1T systems.

In \cite{highspin} the general system with background fields was
studied. It was shown that all possible interactions of a point
particle with background electromagnetic, gravitational and
higher-spin fields in $d$ dimensions emerges from the 2T-physics
worldline theory in Eq.(\ref{free}). The general Sp$(2,R)$
algebraic relations of the $Q_{ij}\left( X,P\right) $ govern the
interactions, and determine equations that the background fields
of any spin must obey. The constraints were solved for a certain
2T to 1T holographic image which describes a relativistic particle
interacting with background fields of any spin in $(d-1)+1$
dimensions. Two disconnected branches of solutions exist, which
seem to have a correspondence with massless states in string
theory, one containing low spins in the zero Regge slope limit,
and the other containing high spins in the infinite Regge slope
limit.

The same kind of holography/duality phenomena that exist in the free case
should, in principle, be expected in the presence of background fields. This
includes holographic capture of the $d+2$ dimensional dynamics in various
forms in $d$ dimensions, duality relations (analogs of same Casimir, and
other related (dual) quantities) among many 1T dynamical systems which have
the same background fields in $d+2$ dimensions, and hidden symmetries in $d$
dimensions which become manifest in $d+2$ dimensions. Such higher symmetries
include global, local, and reparametrization symmetries inherited from $d+2$
dimensions. In principle it is possible to construct numerous duality
relationships as ``experimental'' evidence of the underlying higher
dimensional structure. There is much detail of this type yet to be explored
in the worldline theory. This should be a fruitful area of investigation in
2T-physics.

\subsection{Spin, supersymmetry}

The worldline theory for the spinless particle has been
generalized in several directions. One generalization is to
spinning particles through the use of worldline supersymmetry, in
which case the gauge group is OSp$\left( n|2\right) $ instead of
Sp$\left( 2,R\right) $ \cite{spin2t}$.$ This case has also been
generalized by the inclusion of some background fields \cite
{field2T}, but the most general case analogous to Eq.(\ref{free})
(including all powers of the fermion field) although
straightforward, has not been investigated yet.

Another generalization involves spacetime supersymmetry, which has
been obtained for the free particle (i.e. for $Q_{ij}\rightarrow
q_{ij}$ as in Eq.(\ref{free})) \cite{super2t}\cite{survey2T}. In
this case, in addition to the local Sp$\left( 2,R\right) ,$ there
is local kappa supersymmetry as part of a local supergroup
symmetry. The action is
\begin{equation}
S=\int d\tau \left[ \dot{X}^{M}P_{M}-\frac{1}{2}A^{ij} X_{i}\cdot
X_{j} -\frac{1}{s}Str\left( L\left( \partial _{\tau
}gg^{-1}\right) \right) \right] ,  \label{Lsuper}
\end{equation}
where $g\in G$ is a supergroup element, and $L=L^{MN}\Gamma _{MN}$
is a coupling of the Cartan form $\partial _{\tau }gg^{-1}$ to the
orbital SO$ \left( d,2\right) $ Lorentz generators
$L^{MN}=X^{M}P^{N}-X^{N}P^{M}$ via the spinor representation
$\Gamma _{MN}$ of SO$\left( d,2\right) .$ The supergroup element\
$g$ contains fermions $\Theta $ that are now coupled to phase
space $X^{M},P_{M}$. The constant s is fixed by the dimension of
the spinor representation, and it insures a generalized form of
kappa supersymmetry.

The supergroups OSp$\left( N|4\right) ,$ SU$\left( 2,2|4\right) ,$
F$\left( 4\right) ,$ OSp$\left( 6,2|N\right) ,$ contain SO$\left(
d,2\right) $ in the spinor representation for $d=3,4,5,6$
respectively. The 2T free superparticle of \cite{super2t} based on
these supergroups has a holographic reduction from $d+2$ to $d$
dimensions which produces from Eq.(\ref{Lsuper}) the
superparticle in $d=3,4,5,6,$
\begin{equation}
S=\int d\tau \left[ \dot{x}^{\mu }p_{\mu }-\frac{1}{2}e\left( \tau \right)
p^{2}+\dot{\theta}\gamma ^{\mu }\theta p_{\mu }\right] .
\end{equation}
In these special dimensions the 2T approach of Eq.(\ref{Lsuper})
makes manifest the hidden superconformal symmetry of the
superparticle action which precisely given by the corresponding
supergroup. For other supergroups that contain the bosonic
subgroup SO$\left( d,2\right) $ in the \textit{ spinor
}representation, the 2T supersymmetric model of Eq.(\ref{Lsuper})
includes $p$-brane degrees of freedom. In general, the approach of
\cite {super2t} shows how to formulate any of these systems in
terms of twistors and supertwistors (instead of particle phase
space) by simply choosing gauges, and thus obtaining the spectrum
of the system by using oscillator methods developed a long time
ago \cite{barsgunaydin} (see \cite{lukierski} for a related
twistor approach). For example, OSp$\left( 1|8\right) ,$ which
contains SO$\left( 4,2\right) $= SU$\left( 2,2\right) $ \ (with
$\mathbf{4+4} ^{\ast }$ spinors), is used to construct an action
in $4+2$ dimensions as in Eq.(\ref{Lsuper}); this produces a
holographic picture in $3+1$ dimensions for a superparticle
together with 2-brane degrees of freedom (the 4D superalgebra
containing the maximal 2-brane extension). There is enough gauge
symmetry in the system to remove ghosts associated with timelike
dimensions of the 2-brane. The physical, and unitary, quantum
states of this system correspond to a particle-brane BPS
realization of the supersymmetry OSp$\left( 1|8\right) .$ A
closely related case is the toy M-model in 11+2 dimensions based
on OSp$\left( 1|64\right) $ \cite{super2t}\cite{survey2T}. This
produces a holographic picture that includes the 11-dimensional
2-brane and 5-brane degrees of freedom in addition to the
11-dimensional superparticle phase space (with the maximally
extended superalgebra in 11D). The spectrum of the toy M-model
consists of 2$^{8}$ bosons and $2^{8}$ fermions with the quantum
numbers of the 11D supergravity multiplet, but with a BPS relation
among the brane charges and particle momentum. Another interesting
variation of Eq.(\ref{Lsuper}) is a superparticle model in 10+2
dimensions with SU$\left( 2,2|4\right) $ supersymmetry, in which
the coupling $L$ lives both in SO$\left( 4,2\right) =$SU$\left(
2,2\right) $ and SO$\left( 6\right) =$SU$\left( 4\right) $
\cite{super2t}\cite{survey2T}. This produces an anti-de-Sitter
holographic picture that describes the complete Kaluza-Klein
towers of states that emerge in the AdS$_{5}\times $ S$^{5}$ \
compactification of IIB-supergravity. A generalization of the
latter, including brane degrees of freedom, is achieved by using
OSp$\left( 8|8\right) .$ The methods and partial details of these
constructions are given in \cite{super2t}\cite{survey2T}, and the
full details will appear in the near future.

The generalization of the 2T superparticle with background fields is a
challenging problem that remains to be investigated.

\section{Field theoretic formulation of 2T-physics}

The gauge symmetry Sp$\left( 2,R\right) $ is at the heart of the
worldline formalism and the physical results of 2T-physics. This
is a phase space symmetry as seen from Eq.(\ref{deltaXP}). How
can one implement such a nonlocal gauge symmetry in local field
theory? The answer is naturally found in noncommutative field
theory \cite{NCSp}. In fact, a beautiful and essentially unique
gauge theory formulation of 2T-physics based on noncommutative
u$_{\star }\left( 1,1\right) $ has been obtained  for spinless
particles \cite{NCu11}.

Field theory emerges from the first quantization of the worldline
theory. There is a phase space approach to first quantization
developed in the old days by Weyl-Wigner-Moyal and others
\cite{weyl}-\cite{moyal}. Instead of using wavefunctions in
configuration space $\psi \left( X\right) ,$ this approach uses
wavefunctions in phase space $\phi \left( X,P\right) ,$ which are
equivalent to functions of operators $X,P$ by the Weyl
correspondence. The correct quantum results are produced provided
the phase space wavefunctions are always multiplied with each
other using the noncommutative Moyal star product
\begin{equation}
\left( \phi _{1}\star \phi _{2}\right) \left( X,P\right) =\left.
\exp \left( \frac{i\ \hbar }{2}\frac{\partial }{\partial
X^{M}}\frac{\partial }{\partial \tilde{P}_{M}}-\frac{i\
\hbar}{2}\frac{\partial }{\partial P_{M}}\frac{
\partial }{\partial \tilde{X}^{M}}\right) \phi _{1}\left( X,P\right) \phi
_{2}\left( \tilde{X},\tilde{P}\right) \right| _{X=\tilde{X},\,P=\tilde{P}}.
\end{equation}
Then noncommutative field theory becomes a natural setting for
implementing the local symmetries of 2T-physics.

\subsection{Noncommutative fields from first quantization}

A noncommutative field theory in phase space introduced recently \cite{NCSp}
confirmed the worldline as well as the configuration space field theory \cite
{field2T} results of 2T-physics, and suggested some far reaching insights.
As shown in \cite{NCSp}, first quantization of the worldline theory is
described by the noncommutative field equations
\begin{eqnarray}
\left[ Q_{ij},Q_{kl}\right] _{\star } &=& i\ \hbar \left(
\varepsilon _{jk}Q_{il}+\varepsilon _{ik}Q_{jl}+\varepsilon
_{jl}Q_{ik}+\varepsilon
_{il}Q_{jk}\right) ,  \label{sp2} \\
Q_{ij}\star \varphi &=&0.  \label{matter}
\end{eqnarray}
where the Moyal star product appears in all products. The first
equation is the Sp$\left( 2,R\right) $ commutation relations which
promote the Poisson brackets relations of the worldline theory to
commutators to all orders of $ \hbar $. According to the Weyl
correspondence, we may think of $\varphi $ as an operator in
Hilbert space $\varphi \sim |\psi \rangle \langle \psi |$ (the
equations have a local u$_\star (1)$ symmetry applied on $\varphi$
which allows it to take this special form). The $\varphi $
equations are equivalent to sp$\left( 2,R\right) $ singlet
conditions in Hilbert space, $Q_{ij}|\psi \rangle =0,$ whose
solutions are physical states that are gauge invariant under
sp$\left( 2,R\right) .$ These equations were explicitly solved in
several stages in \cite{field2T}\cite {highspin}\cite{NCSp}. The
solution space is non-empty and is unitary only when spacetime has
precisely two timelike dimensions, no less and no more
\cite{NCSp}. The solution space of these equations confirm the
same physical picture conveyed by the worldline theory, including
the holography/duality and hidden higher dimensional symmetries,
but now in a field theoretical setting \cite{field2T}\cite{NCSp}.
Up to canonical transformations of $ \left( X,P\right) $ the
general solution of Eq.(\ref{sp2}) is given by
\begin{eqnarray}
Q_{11} &=&X^{M}X^{N}\eta _{MN},\quad Q_{12}=X^{M}
P_{M} \label{bkg1} \\
Q_{22} &=&G_{0}\left( X\right) +G_{2}^{MN}\left( X\right) \left( P+A\left(
X\right) \right) _{M}\left( P+A\left( X\right) \right) _{N}  \label{bkg2} \\
&&+\sum_{s=3}^{\infty }G_{s}^{M_{1}\cdots M_{s}}\left( X\right) \,\,\left(
P+A\left( X\right) \right) _{M_{1}}\cdots \left( P+A\left( X\right) \right)
_{M_{s}}  \label{bkg3}
\end{eqnarray}
where $\eta ^{MN}$ is the flat metric in $d+2$ dimensions,
$A_{M}\left( X\right) $ is the Maxwell gauge potential,
$G_{0}\left( X\right) $ is a dilaton, $G_{2}^{MN}\left( X\right)
=\eta ^{MN}+h^{MN}\left( X\right) $ is the gravitational metric,
and the symmetric tensors $\left( G_{s}\left( X\right) \right)
^{M_{1}\cdots M_{s}}$ for $s\geq 3$ are high spin gauge fields.
The sp$\left( 2,R\right) $ closure condition in Eq.(\ref{Qalg})
requires these fields to be homogeneous of degree $(s-2)$ and to
be orthogonal to $X^{M}$
\begin{equation}
X\cdot \partial A_M=-A_M, \quad X\cdot \partial G_{s}=\left(
s-2\right) G_{s},\quad
X^{M}A_{M}=X_{M_{1}}h_{2}^{M_{1}M_{2}}=X_{M_{1}}G_{s\geq
3}^{M_{1}\cdots M_{s}}=0,  \label{hologr}
\end{equation}
There is remaining canonical symmetry which, when expanded in
powers of $P,$ contains the gauge transformation parameters (in
two less dimensions) for all of these background gauge fields
\cite{highspin}. In particular, the Maxwell and Einstein local
symmetries in $d$ dimensions are local symmetries of these $d+2$
dimensional equations which constrain the $d+2$ dimensional
fields to be effectively $d$ dimensional fields (modulo gauge
degrees of freedom that decouple). Thus background fields
$A,G_{0},G_{2},G_{s\geq 3}$ determine all other background fields
$\left( f_{ij}\left( X\right) \right) ^{M_{1}\cdots M_{s}}$ up to
canonical transformations$.$ The solution of the $d+2$ dimensional
equations (\ref {hologr}) is given in \cite{highspin} in terms of
$d$ dimensional background fields for Maxwell $A_{\mu }\left(
x\right) ,$ dilaton $g\left( x\right) ,$ metric $g_{\mu \nu
}\left( x\right) $ and higher spin fields $g^{\mu _{1}\cdots \mu
_{s}}\left( x\right) .$

The solution to the matter field equation is given by a Wigner
distribution function constructed by Fourier transform from a
wavefunctions $ \psi \left( X_{1}\right) =<X_{1}|\psi >$ in
configuration space
\begin{eqnarray}
\varphi \left( X,P\right) &=&\int d^{D}Y\,\psi \left( X\right) \star
e^{-iY^{M}P_{M}}\star \psi ^{\ast }\left( X\right) \\
&=&\int d^{D}Y\,\,\,\psi \left( X-\frac{Y}{2}\right)
\,\,e^{-iY^{M}P_{M}}\,\,\psi ^{\ast }\left( X+\frac{Y}{2}\right) ,
\end{eqnarray}
where the wavefunction satisfies $Q_{ij}|\psi \rangle =0.$ The
sp$\left( 2,R\right) $ gauge invariant solution space
\cite{field2T} of this equation is non-empty and has positive norm
wavefunctions only when there are two timelike dimensions. The
complete set of solutions $\psi _{n}\left( X\right) $ in $d+2$
dimensional configuration space is holographically given
explicitly in 1T spacetime in terms of a complete set of
wavefunctions in $d$ dimensional configuration space. Thus, the
equations (\ref{sp2},\ref{matter} ) correctly represent the 1T
physics of a particle in $d$ dimensions interacting with
background gauge fields, including the Maxwell, Einstein, and high
spin fields \cite{NCSp}.

Using the complete set of physical states, $\psi _{n}\left(
X\right) ,$ one may construct a complete set of physical fields
$\varphi _{m}^{\,\,n}\left( X,P\right) $ in noncommutative 2T
quantum phase space that correspond to $ \varphi _{m}^{\,\,n}\sim
|\psi _{m}\rangle \langle \chi _{n}|$. It can be shown explicitly
that in noncommutative space these complete set of physical fields
satisfy a closed algebra \cite{NCSp}
\begin{equation}
\varphi _{n_{1}}^{\,\,m_{1}}\star \left( \varphi ^{\dagger }\right)
_{m_{2}}^{\,\,n_{2}}\star \varphi _{n_{3}}^{\,\,m_{3}}=\delta
_{\,\,m_{2}}^{m_{1}}\,\delta _{\,\,n_{3}}^{n_{2}}\,\varphi
_{n_{1}}^{\,\,m_{3}},.
\end{equation}
The positive norm (unitarity) of the physical states is captured
by the $ \delta _{nk}$ on the right side.

\subsection{u$_{\star }\left( 1,1\right) $ gauge principle and interactions}

The goal of the field theory approach is to find a field theory,
and appropriate gauge principles, from which the free
Eqs.(\ref{sp2},\ref{matter} ) follow as classical field equations
of motion, much in the same way that
the Klein-Gordon field theory arises from satisfying $\tau $%
-reparametrization constraints ($p^{2}=0$), or string field theory emerges
from satisfying Virasoro constraints, etc. The field theory approach,
combined with gauge principles is expected to provide non-linear field
interactions in 2T-physics.

The desired fundamental gauge symmetry principle that fulfill these goals is
based on noncommutative u$_{\star }\left( 1,1\right) $ in phase space \cite
{NCu11}. There is no non-commutative su$\left( 1,1\right) $ without the
extra u$\left( 1\right) $ in noncommutative space, and therefore to include
sp$\left( 2,R\right) $=su$\left( 1,1\right) $ one must take u$_{\star
}\left( 1,1\right) $ as the smallest candidate symmetry (a smaller candidate
sp$_{\star }\left( 2,R\right) $ \cite{ncOn} which also has a u$_{\star
}\left( 1\right) ,$ is eliminated on other grounds \cite{NCu11}). The
apparently extra noncommutative u$_{\star }\left( 1\right) $ is related to
canonical transformations and plays an important role in the overall scheme.

The u$_{\star }\left( 1,1\right) $ gauge principle completes the formalism
of \cite{NCSp} into an elegant and concise theory which beautifully
describes 2T-physics in field theory in $d+2$ dimensions. The resulting
theory has deep connections to standard $d$ dimensional gauge field
theories, gravity and the theory of high spin fields. There is also a finite
matrix formulation of the theory in terms of u$\left( N,N\right) $ matrices,
such that the $N\rightarrow \infty $ limit becomes the u$_{\star }\left(
1,1\right) $ gauge theory.

The 4 noncommutative parameters of u$_{\star }\left( 1,1\right) $ can be
written in the form of a 2$\times $2 matrix, $\Omega _{ij}=\omega
_{ij}+i\omega _{0}\varepsilon _{ij},$ whose symmetric part $\omega
_{ij}\left( X,P\right) $ becomes sp$\left( 2,R\right) $ when it is global,
while its antisymmetric part generates the local subgroup u$_{\star }\left(
1\right) $ with local parameter $\omega _{0}\left( X_{1},X_{2}\right) .$ The
indices are raised with the sp$\left( 2,R\right) $ metric $\varepsilon
^{ij}, $ therefore in matrix form we have
\begin{equation}
\Omega _{k}^{\,\,l}=\omega _{k}^{\,\,l}-i\omega _{0}\delta
_{k}^{\,\,l}=\left(
\begin{array}{cc}
\omega _{12}-i\omega _{0} & \omega _{22} \\
-\omega _{11} & -\omega _{12}-i\omega _{0}
\end{array}
\right)  \label{omega}
\end{equation}
This matrix satisfies the following hermiticity conditions, $\Omega
^{\dagger }=\varepsilon \Omega \varepsilon .$ Such matrices close under
matrix-star commutators to form u$_{\star }\left( 1,1\right) .$

We introduce a 2$\times $2 matrix $\mathcal{J}_{ij}=J_{ij}+iJ_{0}\varepsilon
_{ij}$ that parallels the form of the parameters $\Omega _{ij}.$ There will
be a close relation between the fields $J_{ij}\left( X,P\right) $ and $%
Q_{ij} $ as we will see soon. When one of the indices is raised,
the matrix $ \mathcal{J}$ takes the form
\begin{equation}
\mathcal{J}_{i}^{\,\,\,\,\,j}=\left(
\begin{array}{cc}
J_{12}-iJ_{0} & J_{22} \\
-J_{11} & -J_{12}-iJ_{0}
\end{array}
\right)  \label{J}
\end{equation}
Next we consider matter fields that transform under the
noncommutative group U$_{\star }^{L}\left( 1,1\right) \times
$U$_{\star }^{R}\left( 1,1\right) .$ In this notation
$\mathcal{J}$ transforms as the adjoint under U$_{\star
}^{L}\left( 1,1\right) $ and is a singlet under U$_{\star
}^{R}\left( 1,1\right) ,$ thus it is in the $\left( 1,0\right) $
representation, which means its gauge transformations are defined
by the matrix-star products in the form $\delta
\mathcal{J}=\mathcal{J}\star \Omega ^{L}-\Omega ^{L}\star
\mathcal{J}$. \ For the matter field we take the $\left(
\frac{1}{2},\frac{1 }{2}\right) $ representation given by a
2$\times 2$ complex matrix $\Phi _{i}^{\,\alpha }\left(
X_{1},X_{2}\right) .$ This field is equivalent to a complex
symmetric tensor $Z_{ij}$ and a complex scalar $\varphi .$ We
define $\bar{\Phi}=\varepsilon \Phi ^{\dagger }\varepsilon .$ The
U$_{\star }^{L}\left( 1,1\right) \times $U$_{\star }^{R}\left(
1,1\right) $ transformation rules for this field are $\delta \Phi
=-\Omega ^{L}\star \Phi +\Phi \star \Omega ^{R},$ where $\Omega
^{L},\Omega ^{R}$ are the infinitesimal parameters for U$_{\star
}^{L}\left( 1,1\right) $ $\times $U$ _{\star }^{R}\left(
1,1\right) .$

We now construct an action that will give the noncommutative field
theory equations (\ref{sp2},\ref{matter}) in a linearized
approximation and provide unique interactions in its full form.
The action has a resemblance to the Chern-Simons type action
introduced in \cite{NCSp}, but now there is one more field,
$J_{0},$ and the couplings among the fields obey a higher gauge
symmetry
\begin{equation}
S_{J,\Phi }=\int d^{2D}X\,Tr\left( -\frac{i}{3}\mathcal{J}\star
\mathcal{J} \star \mathcal{J}-\mathcal{J}\star
\mathcal{J}\mathbf{+\,}i\mathcal{J}\star \Phi \star
\bar{\Phi}-V_{\star }\left( \Phi \star \bar{\Phi}\right) \right) .
\label{SZ}
\end{equation}
The invariance under the local U$_{\star }^{L}\left( 1,1\right)
\times $U$ _{\star }^{R}\left( 1,1\right) $ transformations is
evident. $V\left( u\right) $ is a potential function with argument
$u$=$\Phi \star \bar{\Phi}$.

The form of this action is unique as long as the maximum power of
$\mathcal{J }$ is 3. We have not imposed any conditions on powers
of $\Phi $ or interactions between $\mathcal{J},\Phi ,$ other than
obeying the symmetries. A possible linear term in $\mathcal{J}$
can be eliminated by shifting $\mathcal{J}$ by a constant, while
the relative coefficients in the action are all absorbed into a
renormalization of $\mathcal{J},\Phi $. A term of the form
$Tr\left( \mathcal{J}\star \mathcal{J}\star f\left( \Phi \star
\bar{\Phi}\right) \right) $ that is allowed by the gauge
symmetries can be eliminated by shifting $\mathcal{J}\rightarrow
\left( \mathcal{J}-\frac{1}{3}f\left( \Phi \star \bar{\Phi}\right)
\right) .$ This changes the term $\mathbf{\,}i\bar{ \Phi}\star
\mathcal{J}\star \Phi $ by replacing it with interactions of $
\mathcal{J}$ with any function of $\bar{\Phi},\Phi $ that
preserves the gauge symmetries$.$ However, one can do field
redefinitions to define a new $ \Phi $ so that the interaction
with the linear $\mathcal{J}$ is rewritten as given, thus shifting
all complications to the function $V_{\star }\left(
\bar{\Phi}\star \Phi \right) $. When the maximum power of
$\mathcal{J}$ is cubic we have the correct link to the first
quantized worldline theory. Therefore, with the only assumption
being the cubic restriction on $\mathcal{J},$ this action explains
the first quantized worldline theory, and generalizes it to an
interacting theory based purely on a gauge principle.

The equations of motion are
\begin{equation}
\mathcal{J}\star \mathcal{J}\mathbf{-}2i\mathcal{J}\mathbf{-}\Phi
\star \bar{ \Phi}\mathbf{=}0,\quad \left( \mathcal{J}+iV^{\prime
}\right) \star \Phi =0. \label{generaleq}
\end{equation}
where $V^{\prime }\left( u\right) =\partial V/\partial u$. It is shown in
\cite{NCu11} that one can choose gauges for $\Phi _{i}^{\,\alpha }=\varphi
\delta _{i}^{\,\alpha }$ to simplify these equations, such that $J_{0}$ is
fully solved
\begin{equation}
J_{0}=-1+\frac{V^{\prime }\left( -\left| \lambda \right|
^{2}\right) }{ \left| \lambda \right| ^{2}}\varphi \star \varphi
^{\dagger }+\left( 1-\frac{ 1}{2}J_{ij}\star J^{ij}\right) ^{1/2}.
\label{Jo}
\end{equation}
while the remaining fields satisfy
\begin{equation}
J_{ij}\star \varphi =0,\quad \varphi \varphi ^{\dagger }\varphi =\left|
\lambda \right| \varphi ,\quad 0=\left( 1+V^{\prime }\left( -\left| \lambda
\right| ^{2}\right) \right) ^{2}-1-\left| \lambda \right| ^{2}.
\end{equation}
and
\begin{eqnarray}
\left[ J_{11},J_{12}\right] _{\star } &=&i\left\{ J_{11},\sqrt{1-C_{2}\left(
J\right) }\right\} _{\star }  \label{s1} \\
\left[ J_{11},J_{22}\right] _{\star } &=&2i\left\{ J_{12},\sqrt{%
1-C_{2}\left( J\right) }\right\} _{\star }  \label{s2} \\
\left[ J_{11},J_{22}\right] _{\star } &=&i\left\{ J_{22},\sqrt{1-C_{2}\left(
J\right) }\right\} _{\star }  \label{s3}
\end{eqnarray}
where the expression
\begin{equation}
C_{2}\left( J\right) =\frac{1}{2}J_{kl}\star J^{kl}=\frac{1}{2}J_{11}\star
J_{22}+\frac{1}{2}J_{22}\star J_{11}-J_{12}\star J_{12}
\end{equation}
looks like a Casimir operator. But this algebra is not a Lie algebra, and in
general one cannot show that $C_{2}\left( J\right) $ commutes with $J_{ij}.$
However, assuming no anomalies in the associativity of the star product, the
Jacobi identities $\left[ J_{11},\left[ J_{12},J_{22}\right] _{\star }\right]
_{\star }+cyclic=0$ require
\begin{equation}
\left[ J_{ij},\left[ J^{ij},\sqrt{1-C_{2}\left( J\right) }\right]
_{\star } \right] _{\star }=0,  \label{assoc}
\end{equation}
but generally this is a weaker condition than the vanishing of $\left[
J_{ij},C_{2}\left( J\right) \right] _{\star }.$

To understand the content of the nonlinear gauge field equations
(\ref{s1}- \ref{s3}) we setup a perturbative expansion around a
background solution
\begin{equation}
J_{ij}=J_{ij}^{\left( 0\right) }+gJ_{ij}^{\left( 1\right)
}+g^{2}J_{ij}^{\left( 2\right) }+\cdots
\end{equation}
such that $J_{ij}^{\left( 0\right) }$ is an exact solution, and
then analyze the full equation perturbatively in powers of $g.$
For the exact background solution we assume that
$\frac{1}{2}J_{kl}^{\left( 0\right) }\star J^{(0)kl}$ commutes
with $J_{ij}^{\left( 0\right) },$ therefore the background
solution satisfies a Lie algebra. Then we can write the exact
background solution to Eqs.(\ref{s1}-\ref{s3}) in the form
\begin{equation}
J_{ij}^{\left( 0\right) }=Q_{ij}\star \frac{1}{\sqrt{1+\frac{1}{2}
Q_{kl}\star Q^{kl}}}  \label{JQ}
\end{equation}
where $Q_{ij}$ satisfies the sp$\left( 2,R\right) $ algebra with
the normalization of Eq.(\ref{sp2})
\begin{equation}
\left[ Q_{11},Q_{12}\right] _{\star }=2iQ_{11},\quad \left[
Q_{11},Q_{22} \right] _{\star }=4iQ_{12},\quad \left[
Q_{12},Q_{22}\right] _{\star }=2iQ_{22},  \label{Qalg}
\end{equation}
and $\frac{1}{2}Q_{kl}\star Q^{kl}$ is a Casimir operator that
commutes with all $Q_{ij}$ that satisfies the sp$\left( 2,R\right)
$ algebra. Then the background $Q_{ij}\left( X_{1},X_{2}\right) $
has the form of Eqs.(\ref{bkg1} -\ref{hologr}) up to a u$_{\star
}\left( 1\right) $ subgroup gauge symmetry. The square root is
understood as a power series involving the star products and can
be multiplied on either side of $Q_{ij}$ since it commutes with
the Casimir operator. For such a background, the matter field
equations reduce to
\begin{equation}
Q_{ij}\star \varphi =0.  \label{Qphi}
\end{equation}

Summarizing, we have shown that our action $S_{J,\Phi }$ has
yielded precisely what we had hoped for. The linearized equations
of motion (0$^{th}$ power in $g$) in Eqs.(\ref{Qalg},\ref{Qphi})
are exactly those required by the first quantization of the
worldline theory as given by Eqs.(\ref{sp2}, \ref{matter}). There
remains to understand the propagation and self interactions of the
fluctuations of the gauge fields $gJ_{ij}^{\left( 1\right)
}+g^{2}J_{ij}^{\left( 2\right) }+\cdots $, which are not included
in Eqs.(\ref{Qalg},\ref{Qphi}). However, the full field theory
includes all the information uniquely, in particular the expansion
of Eqs.(\ref{s1}-\ref {s3}) around the background solution
$J_{ij}^{\left( 0\right) }$ of Eq.(\ref {JQ}) should determine
both the propagation and the interactions of the fluctuations
involving photons, gravitons, and high spin fields. At the
linearized level in lowest order, it is shown in \cite{NCu11} that
these fields satisfy the Klein-Gordon type equation.

\section{Remarks and Projects}

We have learned that we can consistently formulate a worldline
theory as well as a field theory of 2T-physics in $d+2$ dimensions
based on basic gauge principles. The equations, compactly written
in $d+2$ dimensional phase space in the form of
Eq.(\ref{generaleq}), yield a unified description of various
gauge fields in configuration space, including Maxwell, Einstein,
and high spin gauge fields interacting with matter and among
themselves in $ d $ dimensions.

All results follow from the field theory action in Eq.(\ref{SZ}),
which is essentially unique save for the assumption of maximum
cubic power of $ \mathcal{J}$. At this time it is not known what
would be the consequences of relaxing the maximum cubic power of
$\mathcal{J}$.

It appears that our approach provides for the first time an action principle
that should contribute to the resolution of the long studied but unfinished
problem of high spin fields \cite{vasil}\cite{segal}\cite{highspin}\cite
{sezgin}. The nature and detail of the interactions can in principle be
extracted from our $d+2$ dimensional theory. The details of the interactions
remain to be worked out, and a comparison to the perturbative equations in
\cite{vasil} is desirable.

For spinning particles the worldline theory introduced osp$\left( n|2\right)
$ in place of sp$\left( 2,R\right) $ \cite{spin2t}. For the field theory
counterpart, we may guess that the appropriate gauge group for the
supersymmetric noncommutative field theory would be u$_{\star }\left(
n|1,1\right) .$

In the case of spacetime supersymmetry, the worldline theory with
background fields remains to be constructed. We expect this to be
a rather interesting and rewarding exercise, because kappa
supersymmetry is bound to require the background fields to satisfy
dynamical equations of motion, as it does in 1T physics
\cite{wittsusy}. The supersymmetric field equations thus obtained
in $d+2$ dimensions should be rather interesting as they would
include some long sought field theories in $d+2$ dimensions, among
them super Yang-Mills and supergravity theories. Perhaps one may
also attempt directly the spacetime supersymmetrization of the
field theory approach, bypassing the background field formulation
of the worldline theory. Based on the arguments given in
\cite{super2t}\cite{survey2T} and \cite{S-theory} we expect the
supersymmetry osp$\left( 1|64\right) $ to play a crucial role in
the relation of this work to M-theory (osp$\left( 1|64\right) $
has also appeared later in \cite{west}\cite{ferrara}).

The noncommutative field theory can be reformulated as a matrix theory in a
large $N$ limit \cite{NCu11}. It is conceivable that these methods would
lead to a formulation of a covariant version of M(atrix) theory \cite{BFSS}.
Using matrix methods one could relate to the 2T-physics formulation of
strings and branes.

A different formulation of 2T-physics for strings (or p-branes) on
the worldsheet (or worldvolume) was initiated in \cite{string2t}.
Tensionless strings (or p-branes) were described in the 2T
approach. However, tensionful strings did not emerge yet in the
formulation. The systematics of the 2T formulation for strings or
branes is not as well understood as the particle case. The
suspicion is that either the action in \cite{string2t} was
incomplete or the correct gauge choice (analogous to the massive
particle) remains to be found. This is still a challenge.

So far the field theory has been analyzed at the classical level. The action
can now be taken as the starting point for a second quantized approach to
2T-physics. The technical aspects of this are open.

It would be interesting to consider phenomenological applications of the
noncommutative field theory approach of 2T-physics, including spinning
particles, and non-Abelian gauge groups.

\bigskip

{\bf Acknowledgements}
\medskip

I would like to thank the organizers of the Karpacz Winter School,
the ITP and the organizers of the Tohwa Symposium, for their
support and for providing a stimulating and enjoyable environment.
This research was partially supported by the US Department of
Energy under grant number DE-FG03-84ER40168 and by the National
Science Foundation under grant Nos. INT97-24831 and PHY99-07949.

\end{document}